\documentclass{article}
\usepackage{graphicx} 
\usepackage{booktabs}
\usepackage{array}
\usepackage{longtable}
\usepackage{placeins}
\usepackage{algorithm}
\usepackage{algpseudocode}
\usepackage[round,authoryear]{natbib}
\usepackage{url}
\usepackage{amssymb, amsmath, amsthm}
\usepackage{thm-restate}
\usepackage{tikz}
\usetikzlibrary{arrows.meta}
\usepackage[hidelinks]{hyperref}
\newcommand{\edgeflow}{F}
\newcommand{\popdemands}{\mathbf{s}}
\newcommand{\policy}{\tau}

\theoremstyle{definition}

\theoremstyle{plain}

\title{\Large Fragmented Obedient Equilibria in Routing Games}
\author{Zhengyang Wang\\
\small Department of Mathematics\\
\small Department of Statistics\\
\small University of Michigan
\and
Achilleas Anastasopoulos\\
\small Department of Electrical Engineering\\
\small and Computer Science\\
\small University of Michigan}
\date{}

\begin{document}

\maketitle

\begin{abstract}
We study routing games in which competing mediators recommend routes to fixed nonatomic populations. Each mediator chooses a randomized flow allocation to minimize its population's expected travel cost, subject to users having no conditional expected gain above a tolerance $\epsilon$ from disregarding their recommendations. We define a fragmented obedient equilibrium (FOE) by combining user obedience with mediator optimality, and establish existence under a lower-hemicontinuity assumption on the obedient policy correspondences. Under affine edge latencies and independent mediator sampling, we derive an exact reformulation in first and second flow moments that preserves best responses and equilibrium profiles. Every feasible moment pair admits a finitely supported realizing policy. When the network has at most four routes, each mediator's best-response problem is an exact semidefinite program. A numerical illustration on the Braess network shows that relaxing obedience can reduce FOE costs, while gaps relative to the social optimum and optimal obedient pooling can persist.
\end{abstract}

\section{Introduction}
\label{sec:introduction}

\subsection{Motivation}

Imagine two colleagues commuting to the same workplace using different navigation platforms. Each receives a private route recommendation and may follow it even when another route offers a small expected saving. Their choices nevertheless interact: both drivers share a road network, so traffic directed by one platform affects travel times for users of the other. Suppose each platform chooses recommendations to minimize its own users’ expected total travel cost, while ensuring that those users have little incentive to deviate. What routing outcomes can arise when the platforms make these decisions independently and strategically? 

In this paper, we model platforms as mediators serving fixed populations, with no switching between mediators. Each mediator chooses a randomized routing policy to minimize its population's expected total travel cost. Users receive private route recommendations, and all routes remain available for deviation. We assume that users follow a recommendation when their conditional expected saving from deviation is at most a tolerance $\epsilon\geq0$.

We call an equilibrium in this setting a fragmented obedient equilibrium (FOE) at tolerance $\epsilon$. Each mediator chooses a best response among policies that satisfy this obedience requirement against fixed rival policies.

\subsection{Related work}

Selfish routing provides the congestion setting and benchmarks for evaluating these outcomes. Wardrop equilibrium describes traffic in which users choose minimum-cost routes, while the social optimum minimizes total travel cost \citep{wardrop1952,beckmann1956studies}. The price of anarchy quantifies the inefficiency of selfish routing, and Braess' paradox shows that adding a link can increase equilibrium travel costs \citep{roughgarden2007routing}.

To account for users' responses to recommendations, we draw on correlated equilibrium and bounded rationality. Correlated equilibrium describes distributions over private action recommendations that players have no incentive to disobey, conditional on their own recommendation and assuming others follow theirs \citep{aumann1974subjectivity}. Boundedly rational user equilibrium (BRUE) allows users to accept routes whose costs are within a tolerance of the minimum available cost \citep{di2013boundedly}. We apply this tolerance to conditional expected travel costs after a recommendation.

Competition among intermediaries makes recommendation policies strategic choices. \citet{tavafoghi2019strategic} study competing navigation platforms with advertising-based and subscription-based business models, accounting for users' platform choices and incentives to follow route recommendations. \citet{jamroz2026competitive} develops competitive mediator games in which users choose a mediator and commit before recommendations are executed.

For fixed populations, splittable routing games model strategic players allocating traffic while accounting for congestion created by others \citep{orda1993competitive}. \citet{roman2023fighting} study competing planners that minimize their assigned populations' travel costs by managing route information sets; restricting those sets to a single route gives direct control over routing. \citet{geffner2025competition} study repeated routing games in which responses to user and planner deviations can sustain efficient outcomes.

Our model combines fixed-population travel-cost minimization with private recommendations subject to conditional approximate obedience.

\subsection{Contributions and numerical illustration}

We establish existence of FOE under a lower-hemicontinuity assumption on the obedient policy correspondences. Under affine edge latencies, we then derive an exact formulation in the first and second moments of each mediator's group flow. This replaces optimization over probability distributions with finite-dimensional moment problems and preserves both best responses and equilibrium profiles. Every feasible moment pair can be realized by a policy that randomizes over finitely many feasible flow allocations. When the route set contains at most four routes, we further express each mediator's best-response problem as an exact semidefinite program.

We illustrate the framework on a three-route Braess network. For several equal-population partitions, we compare numerically verified FOE costs with the social optimum and optimal obedient pooling at the same tolerance. In this example, increasing the obedience tolerance can reduce computed FOE costs, while positive gaps above the social optimum and optimal obedient pooling can remain.

\section{Model}\label{sec:model}

\subsection{Network}

Let $G=(V,E)$ be a finite directed network with origin $O$ and destination $D$, and let $\mathcal R$ be a nonempty finite set of $O$--$D$ routes (i.e., connected edges starting from $O$ and ending at $D$). Each edge $e\in E$ has a continuous, nonnegative, and nondecreasing latency function $c_e:\mathbb R_+\to\mathbb R_+$.

For an aggregate route-flow vector $f=(f_r)_{r\in\mathcal R}$, the load on edge $e$ is
\begin{equation}
    \edgeflow_e=\sum_{r\ni e}f_r.
\end{equation}
The travel cost of route $r$ is
\begin{equation}
    C_r(f)=\sum_{e\in r}c_e(\edgeflow_e),
\end{equation}
and the total social cost is
\begin{equation}
    SC(f)=\sum_{r\in\mathcal R}f_rC_r(f).
\end{equation}
Observe that routes sharing an edge contribute to the same edge load. Thus $C_r(f)$ depends on the aggregate flow vector, including traffic assigned to other routes.

\subsection{Mediators}
There are $K\in \mathbb N$ mediators, indexed by $[K]=\{1,\dots,K\}$. Mediator $k$ serves a fixed nonatomic population with demand $s_k>0$. The demand vector is $\popdemands=(s_k)_{k\in[K]}$. Users do not choose or switch mediators.

A group-$k$ flow assignment is a vector $f^{(k)}=(f_r^{(k)})_{r\in\mathcal R}$ in
\begin{equation}
    \mathcal F_k
    =
    \left\{
        f^{(k)}\in\mathbb R_+^{|\mathcal R|}:
        \sum_{r\in\mathcal R}f_r^{(k)}=s_k
    \right\}.
\end{equation}
For a profile $f^{(1:K)}=(f^{(1)},\dots,f^{(K)})$, the aggregate flow is
\begin{equation}
    f=\sum_{k\in[K]}f^{(k)}.
\end{equation}
For each mediator $k$, write $\mathcal F_{-k}=\prod_{\ell\neq k}\mathcal F_\ell$.

\subsection{Mediation Policies}

Mediator $k$ chooses a recommendation policy $\policy_k\in\Delta(\mathcal F_k)$, where $\Delta(\mathcal F_k)$ denotes the set of Borel probability measures over $\mathcal F_k$. We equip this policy space with the weak topology. A policy profile is $\policy_{1:K}=(\policy_1,\dots,\policy_K)$, and we write $\policy_{-k}=(\policy_\ell)_{\ell\neq k}$ for the rival policies.
The space of rival policy profiles is
\begin{equation}
    \Phi_{-k}
    =
    \prod_{\ell\neq k}\Delta(\mathcal F_\ell).
\end{equation}

Given $\policy_{1:K}$, each mediator independently samples a group recommendation flow $f^{(k)}\sim\policy_k$. The resulting joint distribution is $\bigotimes_{k\in[K]}\policy_k$. When used as an integration measure, $\policy_{-k}$ denotes the product measure induced by the rival policies.

The network, latency functions, population demands, and policy profile $\policy_{1:K}$ are common knowledge. A user in group $k$ observes only their own recommendation from mediator $k$ and does not observe any mediator's sampled group flow.

The sampled flow $f^{(k)}$ specifies the mass of users in group $k$ who receive each route recommendation: mass $f_r^{(k)}$ is recommended route $r$. We assume that recommendations are assigned anonymously, meaning that, conditional on the sampled flow profile, every user in group $k$ has the same probability of receiving recommendation $r$:
\begin{equation}
    \mathbb P\left(\operatorname{rec}_k=r\mid f^{(1:K)}\right)
    =
    \frac{f_r^{(k)}}{s_k}.
    \label{eq:anonymous-recommendation}
\end{equation}
When users follow their recommendations, $f^{(k)}$ is the resulting route flow of population $k$, and $f=\sum_{k\in[K]}f^{(k)}$ determines network congestion.

\subsection{Mediator Objectives}
Each mediator minimizes the expected total travel cost of its own population. Given candidate policy $\policy_k$ and rival policies $\policy_{-k}$, mediator $k$'s objective is
\begin{equation}
    J_k(\policy_k,\policy_{-k})
    =
    \mathbb E_{\policy_k,\policy_{-k}}
    \left[
        \sum_{r\in\mathcal R}
        f_r^{(k)}C_r(f)
    \right].
    \label{eq:mediator-objective}
\end{equation}
The expectation is taken under $\policy_k\otimes\bigotimes_{\ell\neq k}\policy_\ell$, and route costs are evaluated at the aggregate flow $f=\sum_{\ell\in[K]}f^{(\ell)}$.

Thus mediator $k$ accounts for congestion costs borne by its own users, while its routing decisions also affect the costs borne by other populations. For any policy profile,
\begin{equation}
    \sum_{k\in[K]}J_k(\policy_k,\policy_{-k})
    =
    \mathbb E_{\policy_{1:K}}[SC(f)].
\end{equation}

\subsection{Obedience}
Each mediator minimizes its population's total travel cost, while individual users seek to minimize their own travel costs. These objectives need not coincide: a recommendation that benefits the population may leave a user better off taking another route. Every route in $\mathcal R$ remains available to the user, regardless of the recommendation received.

We introduce a tolerance $\epsilon\geq0$ and require that, conditional on the recommendation received, no alternative route offers an expected travel-cost saving greater than $\epsilon$. Because users are nonatomic, a unilateral route deviation does not change aggregate congestion. A group-$k$ user recommended route $r$ therefore evaluates a deviation to route $q$ at the same aggregate flow $f$.

\begin{restatable}[$\epsilon$-obedience]{definition}{epsilonobediencedefinition}
    \label{def:epsilon-obedience}
Fix $\epsilon\geq0$ and rival policies $\policy_{-k}$. A policy $\policy_k\in\Delta(\mathcal F_k)$ is \emph{$\epsilon$-obedient} against $\policy_{-k}$ if, for every $r,q\in\mathcal R$ with $\mathbb E_{\policy_k}[f_r^{(k)}]>0$,
    \begin{equation}
        \mathbb E_{\policy_k,\policy_{-k}}
        \left[
            C_r(f)-C_q(f)
            \mid\operatorname{rec}_k=r
        \right]
        \leq\epsilon.
        \label{eq:conditional-obedience}
    \end{equation}
\end{restatable}
When $\epsilon=0$, no alternative route has a lower expected travel cost than the recommended route. 
For $\epsilon>0$, the tolerance can be interpreted as a friction in how users respond to mediators' recommendations: a deviation becomes desirable only when its conditional expected saving exceeds $\epsilon$. This captures a form of bounded rationality.

The following general identity lets us express expectations conditional on a recommendation in terms of the sampled flows.

\begin{restatable}[Conditional-expectation identity]{lemma}{recommendationconditioning}
\label{lem:recommendation-conditioning}
Let $X$ be a possibly vector-valued random variable, let $A$ be an event, and write $p(X)=\mathbb P(A\mid X)$. If $\mathbb E[p(X)]>0$, then for any bounded measurable function $h$,
\[
    \mathbb E[h(X)\mid A]
    =\frac{\mathbb E[p(X)h(X)]}{\mathbb E[p(X)]}.
\]
\end{restatable}

Fix a policy profile and take $X=f^{(1:K)}$, $A=\{\operatorname{rec}_k=r\}$, and $h(X)=H(f)$, where $f=\sum_{k\in[K]}f^{(k)}$. 
Under the anonymous recommendation convention, $p(X)=f_r^{(k)}/s_k$. Thus, if $\mathbb E[f_r^{(k)}]>0$, the lemma gives
\begin{equation}
    \mathbb E[H(f)\mid\operatorname{rec}_k=r]
    =
    \frac{\mathbb E[f_r^{(k)}H(f)]}{\mathbb E[f_r^{(k)}]}.
    \label{eq:recommendation-conditioning}
\end{equation}

The proof appears in Appendix~\ref{app:model-proof}.

By \eqref{eq:recommendation-conditioning}, the obedience conditions are equivalently
\begin{equation}
    \mathbb E_{\policy_k,\policy_{-k}}
    \left[
        f_r^{(k)}
        \bigl(C_r(f)-C_q(f)-\epsilon\bigr)
    \right]
    \leq0,
    \qquad r,q\in\mathcal R.
    \label{eq:unconditional-obedience}
\end{equation}
This formulation also covers routes recommended with probability zero: if $\mathbb E[f_r^{(k)}]=0$, then $f_r^{(k)}=0$ almost surely, and the corresponding inequalities hold automatically.

For fixed rival policies, define the obedient policy set
\begin{equation}
    \mathcal O_k^\epsilon(\policy_{-k})
    =
    \left\{
        \policy_k\in\Delta(\mathcal F_k):
        \policy_k
        \text{ satisfies \eqref{eq:unconditional-obedience}}
    \right\}.
    \label{eq:obedient-policy-set}
\end{equation}
This is mediator $k$'s feasible set. It depends on rival policies because those policies determine the congestion against which users evaluate their deviations.

\subsection{Fragmented Obedient Equilibrium (FOE)}
Mediator $k$'s obedient best-response correspondence is
\begin{equation}
    BR_k^\epsilon(\policy_{-k})
    =
    \operatorname*{arg\,min}_{
        \policy_k\in\mathcal O_k^\epsilon(\policy_{-k})
    }
    J_k(\policy_k,\policy_{-k}).
    \label{eq:obedient-best-response}
\end{equation}

\begin{restatable}[Fragmented obedient equilibrium]{definition}{foedefinition}
    \label{def:foe}
Fix $\epsilon\geq0$. A policy profile $\policy_{1:K}^\star$ is a \emph{fragmented obedient equilibrium (FOE) at tolerance $\epsilon$} if
    \begin{equation}
        \policy_k^\star\in BR_k^\epsilon(\policy_{-k}^\star),
        \qquad k\in[K].
        \label{eq:foe}
    \end{equation}
We denote the set of such profiles by $\operatorname{FOE}_K(\epsilon)$.
\end{restatable}
At equilibrium, every mediator's policy is $\epsilon$-obedient for its users, and no mediator can reduce its population's expected travel cost by unilaterally choosing another policy that remains $\epsilon$-obedient against the fixed rival policies.

\section{Existence of FOE}\label{sec:existence}

As the rival policy profile $\policy_{-k}$ varies, the set $\mathcal O_k^\epsilon(\policy_{-k})$ defines mediator $k$'s obedient-policy correspondence. We establish existence of an FOE at tolerance $\epsilon$ under a lower-hemicontinuity assumption on these correspondences. The existence proof has three steps. First, we establish joint continuity of each mediator's objective and obedience-constraint expectations in its own policy and its rivals' policies. Second, we show that the obedient-policy sets are nonempty, compact, and convex; their upper hemicontinuity, together with the lower-hemicontinuity assumption, makes the feasible correspondences continuous. Third, Berge's maximum theorem gives upper-hemicontinuous best-response correspondences with nonempty compact convex values. The fixed-point theorem of Kakutani, Fan, and Glicksberg then gives an equilibrium.

The closed-graph property ensures that whenever $\policy_{-k}^n\to\policy_{-k}$ and $\policy_k^n\to\policy_k$ with $\policy_k^n\in\mathcal O_k^\epsilon(\policy_{-k}^n)$ for every $n$, the limiting policy satisfies $\policy_k\in\mathcal O_k^\epsilon(\policy_{-k})$.

This property preserves obedience along convergent sequences. Lower hemicontinuity additionally requires that each obedient policy can be approximated by obedient policies against nearby rival profiles. We impose the following assumption:
\begin{restatable}[Lower hemicontinuity of obedient policy sets]{assumption}{obedientpolicylhc}\label{ass:obedient-policy-lhc}
For the fixed tolerance $\epsilon$, the correspondence $\mathcal O_k^\epsilon$ is lower hemicontinuous for every $k\in[K]$. Equivalently, whenever $\policy_{-k}^n\to\policy_{-k}$ and $\policy_k\in\mathcal O_k^\epsilon(\policy_{-k})$, there exists a sequence of policies $\policy_k^n\in\mathcal O_k^\epsilon(\policy_{-k}^n)$ such that $\policy_k^n\to\policy_k$.
\end{restatable}

Together with Lemma~\ref{lem:obedient-policy-sets}, Assumption~\ref{ass:obedient-policy-lhc} makes each obedient policy correspondence continuous.

\begin{restatable}[Existence of fragmented obedient equilibrium]{theorem}{foeexistence}\label{thm:foe-existence}
Under the model assumptions in Section~\ref{sec:model}, fix $\epsilon\geq0$. If Assumption~\ref{ass:obedient-policy-lhc} holds for every mediator, then $\operatorname{FOE}_K(\epsilon)$ is nonempty.
\end{restatable}

The supporting lemmas and proof appear in Appendix~\ref{app:existence-proof}.

\section{Moment Formulation for Affine Latencies}\label{sec:moment}

Computing mediator $k$'s best response requires optimizing over probability distributions on the set of feasible group flows. Such distributions may have infinite support. Although there are finitely many obedience constraints, the decision variable is an entire measure, which standard finite-dimensional solvers cannot handle directly. Restricting flows to a fixed finite grid generally introduces an approximation.

Under affine edge latencies, we show that a mediator's objective and obedience constraints depend on its policy only through the first and second moments of its group flow. This observation turns each mediator's policy best-response problem into an exact finite-dimensional moment problem.

\citet{zhu2021information} use moment methods to optimize obedient routing policies under uncertain network conditions. Here we derive a moment formulation of the best-response problems of competing mediators.

\subsection{Affine Route Costs}
We specialize the model to edge latencies $c_e(w)=a_ew+b_e$, where $a_e,b_e\geq0$. Let $R=|\mathcal R|$ and fix an ordering of the routes. Define the edge--route incidence matrix $A\in\{0,1\}^{|E|\times R}$ by
\begin{equation}
    A_{er}=
    \begin{cases}
        1, & \text{if route }r\text{ uses edge }e,\\
        0, & \text{otherwise}.
    \end{cases}
\end{equation}
The vector of edge loads is $Af$. Writing $a=(a_e)_{e\in E}$ and $b=(b_e)_{e\in E}$, the route-cost vector $C(f):=(C_r(f))_{r\in\mathcal R}$ is
\begin{equation}
    C(f)=A^\top\bigl(\operatorname{diag}(a)Af+b\bigr)=Hf+d.
    \label{eq:moment-affine-costs}
\end{equation}
Here $H:=A^\top\operatorname{diag}(a)A$, whose entry $H_{rj}=\sum_{e\in r\cap j}a_e$ is the sum of congestion slopes on edges shared by routes $r$ and $j$. Similarly, $d:=A^\top b$, whose entry $d_r=\sum_{e\in r}b_e$ is the constant cost along route $r$.

The matrix $H$ is symmetric and positive semidefinite. The social cost is
\begin{equation}
    SC(f)=f^\top C(f)=f^\top Hf+d^\top f.
    \label{eq:moment-flow-social-cost}
\end{equation}
For matrices, we use the inner product $\langle U,V\rangle=\operatorname{tr}(U^\top V)$.

\subsection{Feasible Flow Moments}
We now represent each policy by the first and second moments of its group flow. These moments cannot be chosen arbitrarily: they must come from a probability distribution over feasible flow allocations. Thus, we define the set of realizable moment pairs and show that every pair in this set can be generated by a feasible policy.

Let $\mathbf 1\in\mathbb R^R$ be the vector of ones. For any mass $s>0$, define $\mathcal F(s)=\{g\in\mathbb R_+^R:\mathbf 1^\top g=s\}$, so $\mathcal F_k=\mathcal F(s_k)$. Define the moment map $T_s:\Delta(\mathcal F(s)) \to \mathbb R^R\times\mathbb S^R$ by
\begin{equation}
    T_s(\policy):=(m,Q),\qquad m:=\mathbb E_\policy[g],\qquad Q:=\mathbb E_\policy[gg^\top],
    \label{eq:moment-map}
\end{equation}
where $g\sim\policy$ and $\mathbb S^R$ denotes the space of real symmetric $R\times R$ matrices. In particular, $m_r=\mathbb E_\policy[g_r]$ and $Q_{rj}=\mathbb E_\policy[g_rg_j]$.

\begin{restatable}[Feasible flow moments]{definition}{feasibleflowmoments}\label{def:feasible-flow-moments}
For $s>0$, define
    \begin{equation}
        \mathfrak M_s:=\operatorname{conv}\{(g,gg^\top):g\in\mathcal F(s)\}\subseteq\mathbb R^R\times\mathbb S^R,
        \label{eq:feasible-flow-moments}
    \end{equation}
where $\operatorname{conv}$ denotes finite convex combinations.
\end{restatable}

\begin{restatable}[Exact moment feasibility]{proposition}{momentfeasibility}\label{prop:moment-feasibility}
For every $s>0$, the set $\mathfrak M_s$ is nonempty, compact, and convex, and
    \begin{equation}
        T_s\bigl(\Delta(\mathcal F(s))\bigr)=\mathfrak M_s.
        \label{eq:moment-image}
    \end{equation}
Every pair in $\mathfrak M_s$ can be realized by a policy supported on at most $R(R+1)/2$ feasible flows.
\end{restatable}

The proposition lets us replace a distribution by its moments and still recover a feasible policy, even when the original distribution has infinite support. The recovered policy need not be the original one: different policies can have the same $(m,Q)$. Every realizable pair satisfies the mass equalities $\mathbf1^\top m=s$ and $Q\mathbf1=sm$, and its second-moment matrix is positive semidefinite and entrywise nonnegative. If $m_r=0$, route $r$ is unused almost surely, so $Q_{rj}=0$ for every $j$.

The proof of Proposition~\ref{prop:moment-feasibility} and the derivation of these necessary properties in Lemma~\ref{lem:moment-identities} appear in Appendix~\ref{app:moment-proof}.

\subsection{Objectives and Obedience in Moments}
We now express mediator costs, user obedience, and expected social cost in moment coordinates. These formulas identify the information about policies needed to evaluate mediator decisions and their effects on users.

Fix mediator $k$ and rival policies $\policy_{-k}$. For each rival $\ell\neq k$, let $m^{(\ell)}=\mathbb E_{\policy_\ell}[f^{(\ell)}]$ and define the aggregate rival mean $v_k:=\sum_{\ell\neq k}m^{(\ell)}$, with $v_1=0$ when $K=1$. Let $(m,Q)=T_{s_k}(\policy_k)$ for mediator $k$'s candidate policy. Because the aggregate flow $f_j$ includes both own-group and rival flow, independence across mediators gives
\begin{equation}
    \mathbb E_{\policy_k,\policy_{-k}}[f_r^{(k)}f_j]=Q_{rj}+m_r(v_k)_j.
    \label{eq:moment-cross-flow}
\end{equation}

The first term is a within-group second moment; the second is the product of the own-group mean and the aggregate rival mean.

\begin{restatable}[Moment formulas for objectives, obedience, and social cost]{proposition}{momentcostobedience}\label{prop:moment-cost-obedience}
For a candidate policy $\policy_k$ with moments $(m,Q)$, mediator $k$'s objective is
    \begin{equation}
        J_k(\policy_k,\policy_{-k})=\langle H,Q\rangle+m^\top Hv_k+d^\top m.
        \label{eq:moment-mediator-objective}
    \end{equation}
The policy is $\epsilon$-obedient against $\policy_{-k}$ if and only if, for every pair of distinct routes $r,q\in\mathcal R$,
    \begin{equation}
        \sum_{j\in\mathcal R}(H_{rj}-H_{qj})\bigl(Q_{rj}+m_r(v_k)_j\bigr)+(d_r-d_q-\epsilon)m_r\leq0.
        \label{eq:moment-obedience}
    \end{equation}
At a full policy profile $\policy_{1:K}$, write $(m^{(k)},Q^{(k)})=T_{s_k}(\policy_k)$ and $\bar m=\sum_km^{(k)}$. The expected social cost is
    \begin{equation}
        \mathbb E_{\policy_{1:K}}[SC(f)]=\sum_k\langle H,Q^{(k)}\rangle+2\sum_{k<\ell}(m^{(k)})^\top Hm^{(\ell)}+d^\top\bar m.
        \label{eq:moment-social-cost}
    \end{equation}
\end{restatable}

For fixed rivals, the objective and obedience constraints are linear in the candidate moments $(m,Q)$, and rival policies matter only through $v_k$. Policies with the same moments therefore have the same objective and obedience status against the same rivals. The proof appears in Appendix~\ref{app:moment-proof}. Appendix~\ref{app:moment-cost-consequences} also derives the conditional expected route-cost difference and the decomposition of expected social cost into cost at the mean flow and a covariance contribution.

\subsection{Best Responses and Equilibrium in Moment Space}
The preceding formulas allow each mediator to optimize over feasible moments while holding its rivals' aggregate mean flow fixed. We now define best responses in moment space and show that simultaneous moment best responses correspond exactly to FOE.

For mediator $k$ and an aggregate rival mean $v$ arising from rival policies, define the obedient moment set
\begin{equation}
    \mathfrak O_k^\epsilon(v):=\left\{(m,Q)\in\mathfrak M_{s_k}:\text{\eqref{eq:moment-obedience} holds with }v_k=v\right\}.
    \label{eq:obedient-moment-set}
\end{equation}
Its moment best-response correspondence is
\begin{equation}
    BR_{k,\mathrm{mom}}^\epsilon(v):=\operatorname*{arg\,min}_{(m,Q)\in\mathfrak O_k^\epsilon(v)}\bigl[\langle H,Q\rangle+m^\top Hv+d^\top m\bigr].
    \label{eq:moment-best-response}
\end{equation}

\begin{restatable}[Moment-space FOE]{definition}{mfoedefinition}\label{def:moment-equilibrium}
A moment-space fragmented obedient equilibrium (mFOE) at tolerance $\epsilon$ is a profile $((m^{(k)},Q^{(k)}))_{k\in[K]}\in\prod_k\mathfrak M_{s_k}$ such that, for each $k$, with $v_k=\sum_{\ell\neq k}m^{(\ell)}$,
    \begin{equation}
        (m^{(k)},Q^{(k)})\in BR_{k,\mathrm{mom}}^\epsilon(v_k).
        \label{eq:moment-equilibrium}
    \end{equation}
We denote the set of such profiles by $\operatorname{mFOE}_K(\epsilon)$.
\end{restatable}

\begin{restatable}[Exact moment reformulation]{theorem}{momentequivalence}\label{thm:moment-equivalence}
Under affine edge latencies and independent mediator sampling,
    \[
        \policy_{1:K}\in\operatorname{FOE}_K(\epsilon)
        \quad\Longleftrightarrow\quad
        (T_{s_k}(\policy_k))_{k\in[K]}\in\operatorname{mFOE}_K(\epsilon).
    \]
Every profile in $\operatorname{mFOE}_K(\epsilon)$ can be realized by finitely supported policies.
\end{restatable}

All realizations of the same moment profile have identical expected mediator costs and expected social cost by \eqref{eq:moment-mediator-objective} and \eqref{eq:moment-social-cost}. Under the assumptions of Theorem~\ref{thm:foe-existence}, $\operatorname{mFOE}_K(\epsilon)$ is nonempty. The proof and supporting properties of the moment feasible and best-response sets appear in Appendix~\ref{app:moment-proof}.

\subsection{Exact Semidefinite Best Responses}\label{sec:moment-conic}
We can now turn the moment best-response problem into a form accepted by a standard solver. For networks with at most four suggested routes, positive semidefiniteness, entrywise nonnegativity, and the mass equalities characterize realizable moment pairs exactly. Adding the obedience constraints gives the following semidefinite program.

\begin{restatable}[Exact semidefinite best responses]{theorem}{momentsdp}\label{cor:moment-sdp}
If $R\leq4$, then for fixed rival policies with aggregate mean $v_k$, mediator $k$'s moment best responses are exactly the solutions of the semidefinite program
    \begin{equation}
        \begin{aligned}
            \underset{m\in\mathbb R^R,\ Q\in\mathbb S^R}{\operatorname{minimize}}\quad
            &\langle H,Q\rangle+m^\top Hv_k+d^\top m\\
            \text{subject to}\quad
            &Q\succeq0,\qquad Q_{rj}\geq0\quad\forall r,j,\\
            &Q\mathbf1=s_km,\qquad\mathbf1^\top m=s_k,\\
            &\sum_j(H_{rj}-H_{qj})\bigl(Q_{rj}+m_r(v_k)_j\bigr)\\
            &\qquad+(d_r-d_q-\epsilon)m_r\leq0\quad\forall r\neq q.
        \end{aligned}
        \label{eq:moment-sdp}
    \end{equation}
Every optimal moment pair has a realizing policy that is a best response against these rivals.
\end{restatable}

The proof and the construction of realizing policies appear in Appendix~\ref{app:moment-conic}. For $R>4$, the displayed moment constraints do not guarantee realizability, so the SDP is generally a relaxation of the exact moment best-response problem.

The condition $R\leq4$ need not be restrictive in applications: it limits the number of route alternatives considered in the model, rather than the size or density of the network. The network may be arbitrarily large and dense, provided that $\mathcal R$ contains at most four $O$--$D$ routes treated as legitimate alternatives for both recommendations and user deviations.

\section{Computation and Numerical Illustration}\label{sec:numerical-method}\label{sec:numerical-analysis}

For $R\leq4$, we use the moment SDP to compute and verify candidate equilibrium profiles, then illustrate their expected social costs on the Braess network.

\subsection{Computing and Verifying Equilibrium}

Write $x_k^t=(m_k^t,Q_k^t)$ for mediator $k$'s moments after sweep $t$, and $x^t=(x_k^t)_{k\in[K]}$ for the moment profile. Algorithm~\ref{alg:numerical-procedure} uses sequential updates: each mediator responds to the latest available rival means. After computing a numerical best response $\widehat x_k^t$ from \eqref{eq:moment-sdp}, damping applies
\begin{equation}
    x_k^{t+1}=(1-\lambda)x_k^t+\lambda\widehat x_k^t.
    \label{eq:numerical-damping}
\end{equation}
The illustration uses $\lambda=0.5$ and at most 160 sweeps. Convexity preserves realizability in exact arithmetic. However, each best response is computed against the current rival profile, which changes during subsequent updates. We therefore verify obedience and mediator optimality against the final rival profile rather than infer equilibrium from small updates. Convergence of the procedure is not proved.

\begin{algorithm}[H]
    \caption{Sequential damped best responses and policy verification}\label{alg:numerical-procedure}\label{fig:numerical-procedure}
    \begin{algorithmic}[1]
        \Require $H,d,(s_k),\epsilon$, realizable $x^0$, damping $\lambda\in(0,1]$, limit $T$, numerical tolerances
        \Ensure Numerically verified finitely supported policies or a failure status
        \For{$t=0,\ldots,T$}
            \If{$\textsc{ShouldCheck}(t,x^t)$}
                \If{$\textsc{VerifyMomentEquilibrium}(x^t)$}
                    \State \Return $\textsc{RealizePolicies}(x^t)$
                \EndIf
            \EndIf
            \If{$t=T$}
                \State \Return iteration-limit status
            \EndIf
            \State $x^{t+1}\gets x^t$
            \For{$k=1,\ldots,K$}
                \State $v_k\gets\sum_{\ell\neq k}m_\ell^{t+1}$
                \State $\widehat x_k^t\gets\textsc{SolveBestResponse}(v_k)$
                \State \Return failure if the best-response solve fails its checks
                \State $x_k^{t+1}\gets(1-\lambda)x_k^t+\lambda\widehat x_k^t$
            \EndFor
        \EndFor
    \end{algorithmic}
\end{algorithm}

In Algorithm~\ref{alg:numerical-procedure}, \textsc{ShouldCheck} determines whether to run an equilibrium check at the current sweep. If so, \textsc{VerifyMomentEquilibrium} checks moment feasibility, user obedience, and whether each mediator's moments form a best response to the current rival means, within the numerical tolerances. \textsc{SolveBestResponse} solves \eqref{eq:moment-sdp} against the current rival means. If the moment profile passes verification, \textsc{RealizePolicies} constructs finitely supported policies, checks their reconstruction accuracy, and rechecks obedience and mediator optimality using the reconstructed moments and rival means. It returns the verified policies, or a failure status if construction or verification fails. Appendix~\ref{app:numerics} gives the verification formulas, thresholds, and recovery settings.

\FloatBarrier
\subsection{Illustration on the Braess Network}\label{sec:numerical-design}\label{sec:numerical-plateaus}

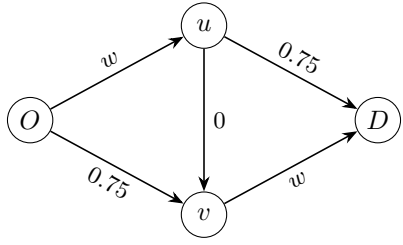
\begin{figure}[H]
    \centering
    \begin{tikzpicture}[
        vertex/.style={circle,draw,minimum size=6mm,inner sep=1pt},
        edge/.style={-{Stealth[length=2mm]},semithick}
    ]
        \node[vertex] (O) at (0,0) {$O$};
        \node[vertex] (u) at (2.3,1.25) {$u$};
        \node[vertex] (v) at (2.3,-1.25) {$v$};
        \node[vertex] (D) at (4.6,0) {$D$};
        \draw[edge] (O) -- node[above,sloped,font=\small] {$w$} (u);
        \draw[edge] (u) -- node[above,sloped,font=\small] {$0.75$} (D);
        \draw[edge] (O) -- node[below,sloped,font=\small] {$0.75$} (v);
        \draw[edge] (v) -- node[below,sloped,font=\small] {$w$} (D);
        \draw[edge] (u) -- node[right,font=\small] {$0$} (v);
    \end{tikzpicture}
    \caption{Braess network used in the illustration. Edge labels give latencies as functions of edge load $w$.}
    \label{fig:numerical-braess-network}
\end{figure}

We illustrate the computation on the three-route Braess network in Figure~\ref{fig:numerical-braess-network}, with unit demand. The routes are $T=(O,u,D)$, $M=(O,u,v,D)$, and $B=(O,v,D)$. Edges $O\to u$ and $v\to D$ have latency $c(w)=w$, edges $u\to D$ and $O\to v$ have constant latency $0.75$, and the middle edge $u\to v$ has zero latency. We choose the constant latency $0.75$ to give an example with distinct social-optimum and Wardrop costs. We consider $K=2,4,8$ equal subpopulations, with $s_k=1/K$, and vary the obedience tolerance while holding the network and each population partition fixed.

The social optimum (SO) minimizes $SC(f)$ over $\mathcal F(1)$. In this network, $f^{\mathrm{SO}}=(0.5,0,0.5)$ in route order $(T,M,B)$: half the population uses the top route and half uses the bottom route, leaving the middle route unused. Its cost is $S_{\mathrm{SO}}:=SC(f^{\mathrm{SO}})=1.25$.
Optimal obedient pooling is the minimum expected social cost achievable by a single mediator serving the full population, subject to user obedience at the same tolerance $\epsilon$; denote its cost by $P(\epsilon)$. Every obedient fragmented profile satisfies the pooling bound below. In particular, for any equilibrium profile $\policy_{1:K}^{\star}\in\operatorname{FOE}_K(\epsilon)$,
\begin{equation}
    S_{\mathrm{SO}}\leq P(\epsilon)\leq\mathbb E_{\policy_{1:K}^{\star}}[SC(f)].
    \label{eq:numerical-pooling-bound}
\end{equation}
Appendix~\ref{app:numerical-benchmarks} provides the justification.

For each computed profile, we report two measures:
\begin{equation}
    \begin{aligned}
        E_{\mathrm{SO}}&:=100\frac{\mathbb E_{\policy_{1:K}^{\star}}[SC(f)]-S_{\mathrm{SO}}}{S_{\mathrm{SO}}},\\
        E_{\mathrm{pool}}&:=100\frac{\mathbb E_{\policy_{1:K}^{\star}}[SC(f)]-P(\epsilon)}{S_{\mathrm{SO}}}.
    \end{aligned}
    \label{eq:numerical-efficiency-metrics}
\end{equation}
Both are expressed as percentages of SO cost. The first measures overall inefficiency; the second compares fragmented routing with a pooled decision-maker facing the same obedience restriction.

A configuration specifies the network, edge latencies, total demand, mediator populations, and obedience tolerance. Here the network, latencies, and unit demand remain fixed; configurations differ in $K\in\{2,4,8\}$, with equal masses $s_k=1/K$, and in $\epsilon$.

For each configuration, we run Algorithm~\ref{alg:numerical-procedure} from two randomized initial policy profiles and one proportional Wardrop profile. In each randomized profile, every mediator draws four route-share vectors from $\operatorname{Dirichlet}(1,1,1)$ and scales them by $s_k$ to obtain feasible allocations. Their mixture probabilities are drawn from $\operatorname{Dirichlet}(1,1,1,1)$. These Dirichlet distributions sample nonnegative shares summing to one uniformly over their respective simplexes. The proportional Wardrop initialization assigns each mediator the deterministic allocation $s_kw$, where $w$ is the aggregate Wardrop flow.

\begin{figure}[tb]
    \centering
    \includegraphics[width=\linewidth]{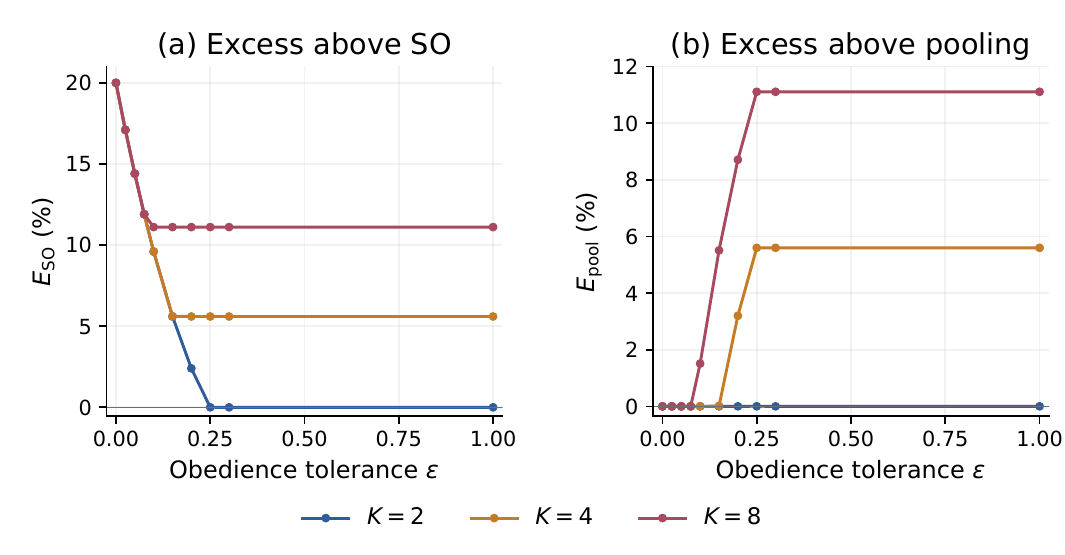}
    \caption{Braess network with equal populations. Left: $E_{\mathrm{SO}}$, excess expected social cost above SO. Right: $E_{\mathrm{pool}}$, excess above optimal obedient pooling at the same $\epsilon$, also divided by SO cost. Lines connect medians across three starts at the sampled tolerances; shaded bands span all three starts. At $\epsilon=1$, obedience is unrestricted.}
    \label{fig:numerical-tolerance}
\end{figure}

After running the algorithm and verifying the final profiles, we take the median of their three final expected social costs and express it through the two metrics in \eqref{eq:numerical-efficiency-metrics}. Figure~\ref{fig:numerical-tolerance} shows $E_{\mathrm{SO}}$ in the left panel and $E_{\mathrm{pool}}$ in the right panel. Curves show the median across the three initializations, with shaded bands spanning their minimum and maximum values. The bands are generally narrow and overlap the median curves. Appendix~\ref{app:numerics} records the random seeds, tolerance grid, and verification settings.

At $\epsilon=0.05$, the computed FOE costs and optimal pooling cost are all approximately $1.43$: excess above SO is about $14.4\%$, while excess above pooling is negligible. As the tolerance increases, costs fall but the population partitions yield different outcomes. At $\epsilon=0.25$, pooling and the two-mediator profiles approach SO, whereas four- and eight-mediator profiles remain near $1.32$ and $1.3889$, respectively. Their excess costs above SO are about $5.6\%$ and $11.1\%$. These gaps persist at $\epsilon=1$, where all obedience constraints are redundant and optimal pooling attains SO. The example therefore shows that relaxing obedience need not eliminate the efficiency gap. The Braess example illustrates how the framework can be used to analyze FOE costs on other networks, but does not establish how these costs vary with the number of mediators or the obedience tolerance in general.

\section{Discussion}\label{sec:discussion}

In a framework where two major factors jointly determine the routing outcomes, users decide whether to follow recommendations, while mediators choose policies to minimize their own populations' travel costs. 
The moment formulation makes this interaction computationally accessible under affine latencies. The Braess illustration demonstrates how the framework can compare fragmented routing with the social optimum and optimal obedient pooling.

In the tested configurations, increasing the obedience tolerance can reduce equilibrium costs, but positive efficiency gaps remain even when obedience no longer constrains policies. 
Thus, removing the obedience restriction need not eliminate the efficiency gap between competing mediators and the social optimum.
These observations motivate further investigation across network structures, population partitions, and equilibrium selection procedures.

A natural extension is to networks with multiple $O$--$D$ pairs, with several mediators serving each pair. Mediators would interact through congestion on shared edges, both within and across $O$--$D$ pairs. If each mediator serves a single pair, its allocation vector would involve only that pair's routes, while travel costs would depend on traffic from all pairs. 
An interesting question is whether the moment formulation can extend to this setting while retaining an exact SDP for each best response when each pair has at most four routes.

Several theoretical and computational questions remain. Equilibrium existence currently relies on lower hemicontinuity of the obedient-policy correspondences, and convergence of the damped best-response procedure is not established. 
Beyond four routes, the doubly nonnegative constraints generally provide a relaxation, so additional conditions or computational methods are needed to guarantee policy realizability. Addressing these questions would broaden the settings in which the framework can be used to analyze FOE.

\section{Conclusion}\label{sec:conclusion}

We introduced fragmented obedient equilibrium for competing routing mediators serving fixed populations and established existence under lower hemicontinuity of the obedient-policy correspondences. For affine edge latencies, the exact moment reformulation preserves best responses and equilibria and permits finitely supported policy realizations; with at most four routes, best responses can be computed by semidefinite programming. Together with the Braess illustration, these results provide a framework for studying routing outcomes under user obedience and competition among mediators.

\clearpage
\bibliographystyle{plainnat-linked}
\bibliography{references}

\clearpage
\appendix
\section{Proof from the Model}\label{app:model-proof}

\recommendationconditioning*

\begin{proof}[Proof of Lemma~\ref{lem:recommendation-conditioning}]
By the law of iterated expectations,
\begin{align*}
    \mathbb P(A)
    &=\mathbb E[\mathbf 1_A] \\
    &=\mathbb E\left[\mathbb E[\mathbf 1_A\mid X]\right] \\
    &=\mathbb E[p(X)]>0.
\end{align*}
Consequently, the conditional expectation is well defined and satisfies
\[
    \mathbb E[h(X)\mid A]
    =\frac{\mathbb E[h(X)\mathbf 1_A]}{\mathbb P(A)}.
\]
To evaluate the numerator, condition on $X$. Since $h(X)$ is measurable with respect to $X$,
\begin{align*}
    \mathbb E[h(X)\mathbf 1_A]
    &=\mathbb E\left[\mathbb E[h(X)\mathbf 1_A\mid X]\right] \\
    &=\mathbb E\left[h(X)\mathbb P(A\mid X)\right] \\
    &=\mathbb E[p(X)h(X)].
\end{align*}
Dividing the numerator by the event probability yields
\[
    \mathbb E[h(X)\mid A]
    =\frac{\mathbb E[p(X)h(X)]}{\mathbb E[p(X)]}.
\]
\end{proof}

\section{Proof of the Existence Theorem}\label{app:existence-proof}

\subsection{Continuity in Policies}
Fix $\epsilon\geq 0$. The space of group-flow profiles is $\prod_{k\in[K]}\mathcal F_k$, and the policy-profile space is $\Phi=\prod_{k\in[K]}\Delta(\mathcal F_k)$. Each mediator's policy space carries the weak topology, and $\Phi$ carries the product weak topology. Thus a sequence of policy profiles converges if and only if each mediator's policy converges weakly.

\begin{restatable}[Compactness and convexity of $\Delta(\mathcal F_k)$]{lemma}{policyspaces}\label{lem:policy-spaces}
For every mediator $k$, the policy space $\Delta(\mathcal F_k)$ is nonempty, compact, metrizable, and convex under the weak topology. The policy-profile space $\Phi$ has the same properties under the product weak topology, and is a subset of a locally convex Hausdorff topological vector space.
\end{restatable}

\begin{proof}
Each $\mathcal F_k$ is a nonempty compact simplex equipped with the Euclidean metric. Theorem~15.11 of \citet{aliprantis2006infinite} therefore implies that $\Delta(\mathcal F_k)$ is compact and metrizable under the weak topology. It is nonempty because it contains the point mass at any flow in $\mathcal F_k$, and it is convex because mixtures of probability measures are probability measures. Since $K$ is finite, these properties also hold for the product space $\Phi$.

Let $\mathcal M_k$ be the vector space of finite signed Borel measures on $\mathcal F_k$, equipped with the topology generated by the seminorms $p_h(\mu)=|\int_{\mathcal F_k}h\,d\mu|$, where $h\in C(\mathcal F_k)$. This topology is locally convex. Since continuous functions separate finite signed Borel measures on the compact metric space $\mathcal F_k$, the seminorm family is separated, so $\mathcal M_k$ is Hausdorff by the Hausdorff criterion.

The induced topology on $\Delta(\mathcal F_k)\subseteq\mathcal M_k$ is the weak topology. Therefore $\Phi$, with the product weak topology, is a subset of the locally convex Hausdorff topological vector space $\prod_{k\in[K]}\mathcal M_k$.
\end{proof}

\begin{restatable}{lemma}{policycontinuity}\label{lem:policy-continuity}
For every $k\in[K]$ and $r,q\in\mathcal R$, the mediator's objective
    \begin{equation*}
        J_k(\policy_k,\policy_{-k}) := 
        \mathbb E_{\policy_k,\policy_{-k}}
        \left[
            \sum_{r\in\mathcal R}
            f_r^{(k)}C_r(f)
        \right]
    \end{equation*}
and the expectation
    \begin{equation*}
        \mathbb E_{\policy_k,\policy_{-k}}\bigl[
            f_r^{(k)}\bigl(C_r(f)-C_q(f)-\epsilon\bigr)
        \bigr]
    \end{equation*}
are jointly continuous in $(\policy_k,\policy_{-k})$ and linear in $\policy_k$ for every fixed $\policy_{-k}$.
\end{restatable}

\begin{proof}
For a group-flow profile $f^{(1:K)}\in\prod_{\ell\in[K]}\mathcal F_\ell$, the aggregate flow is $f=\sum_{\ell\in[K]}f^{(\ell)}$. Aggregation and the route-cost functions are continuous. Therefore both integrands
    \begin{gather*}
        \sum_{r\in\mathcal R}f_r^{(k)}C_r(f),\\
        f_r^{(k)}\bigl(C_r(f)-C_q(f)-\epsilon\bigr)
    \end{gather*}
are continuous on the compact space $\prod_{\ell\in[K]}\mathcal F_\ell$ and are consequently bounded.
    
Suppose $\policy_k^n\to\policy_k$ weakly and $\policy_{-k}^n\to\policy_{-k}$ in the product weak topology. The induced product measures converge weakly on $\prod_{\ell\in[K]}\mathcal F_\ell$. Integration of either bounded continuous function against these measures therefore gives
    \begin{equation*}
        J_k(\policy_k^n,\policy_{-k}^n)
        \longrightarrow J_k(\policy_k,\policy_{-k})
    \end{equation*}
and
    \begin{gather*}
        \mathbb E_{\policy_k^n,\policy_{-k}^n}\bigl[
            f_r^{(k)}\bigl(C_r(f)-C_q(f)-\epsilon\bigr)
        \bigr]\\
        \longrightarrow
        \mathbb E_{\policy_k,\policy_{-k}}\bigl[
            f_r^{(k)}\bigl(C_r(f)-C_q(f)-\epsilon\bigr)
        \bigr].
    \end{gather*}
This proves continuity. With rival policies fixed, linearity in $\policy_k$ follows from linearity of integration.
    
\end{proof}

\subsection{Obedient Policy Sets}
\begin{restatable}[Properties of obedient policy sets]{lemma}{obedientpolicysets}\label{lem:obedient-policy-sets}
\label{lem:obedient-policy-values}
\label{lem:obedient-policy-uhc}
For every mediator $k$, the correspondence $\mathcal O_k^\epsilon:\Phi_{-k}\rightrightarrows\Delta(\mathcal F_k)$ has nonempty compact convex values, has a closed graph, and is upper hemicontinuous.
\end{restatable}
\begin{proof}
    \textit{Convexity.} Fix $k$ and $\policy_{-k}$. Each obedience constraint in \eqref{eq:unconditional-obedience} is linear in $\policy_k$, so any convex combination of two obedient policies remains obedient. Hence $\mathcal O_k^\epsilon(\policy_{-k})$ is convex.

    \textit{Compactness.} To prove closedness, take any sequence $\policy_k^n\in\mathcal O_k^\epsilon(\policy_{-k})$ converging weakly to some $\policy_k\in\Delta(\mathcal F_k)$. By Lemma~\ref{lem:policy-continuity}, for every $r,q\in\mathcal R$,
    \begin{gather*}
        \mathbb E_{\policy_k,\policy_{-k}}\left[
            f_r^{(k)}\bigl(C_r(f)-C_q(f)-\epsilon\bigr)
        \right]\\
        =
        \lim_{n\to\infty}
        \mathbb E_{\policy_k^n,\policy_{-k}}\left[
            f_r^{(k)}\bigl(C_r(f)-C_q(f)-\epsilon\bigr)
        \right]
        \leq 0.
    \end{gather*}
The inequality holds because every $\policy_k^n$ is obedient. Thus $\policy_k$ also satisfies every obedience constraint and belongs to $\mathcal O_k^\epsilon(\policy_{-k})$. Since $\Delta(\mathcal F_k)$ is metrizable, this proves that $\mathcal O_k^\epsilon(\policy_{-k})$ is closed. It is therefore compact as a closed subset of the compact policy space $\Delta(\mathcal F_k)$ established in Lemma~\ref{lem:policy-spaces}.

    \textit{Nonemptiness.} To prove nonemptiness, fix $\policy_{-k}$ and define the expected route costs
    \begin{equation}
        \bar C_r(g)
        :=\mathbb E_{\policy_{-k}}\left[
            C_r\left(g+\sum_{\ell\neq k}f^{(\ell)}\right)
        \right],
        \qquad g\in\mathcal F_k.
        \label{eq:expected-rival-route-cost}
    \end{equation}
These functions are continuous because route costs are continuous on the compact feasible flow space. Define
    \[
        W(g)
        :=\operatorname*{arg\,min}_{z\in\mathcal F_k}
        \sum_{r\in\mathcal R}z_r\bar C_r(g).
    \]
The route costs are evaluated at $g$ and held fixed while optimizing over $z$. The objective is jointly continuous in $(g,z)$, and the feasible set $\mathcal F_k$ is nonempty, compact, and independent of $g$. The minimization form of Berge's maximum theorem therefore implies that $W$ has nonempty compact values and is upper hemicontinuous \citep{aliprantis2006infinite}. Its values are convex because the objective is linear in $z$. Since $\mathcal F_k$ is also convex, Kakutani's fixed-point theorem yields $g^\star\in W(g^\star)$ \citep{delatte2022kakutani}.
    
At this fixed point,
    \[
        g_r^\star>0
        \quad\Longrightarrow\quad
        \bar C_r(g^\star)\leq\bar C_q(g^\star)
        \quad\text{for every }q\in\mathcal R;
    \]
otherwise, shifting mass from $r$ to a cheaper route would contradict $g^\star\in W(g^\star)$. Let $\delta_{g^\star}$ be the policy that selects $g^\star$ with probability one. For every $r,q\in\mathcal R$,
    \begin{align*}
        &\mathbb E_{\delta_{g^\star},\policy_{-k}}\left[
            f_r^{(k)}\bigl(C_r(f)-C_q(f)-\epsilon\bigr)
        \right]\\
        &\qquad=
        g_r^\star\bigl(\bar C_r(g^\star)-\bar C_q(g^\star)-\epsilon\bigr)
        \leq0.
    \end{align*}
The inequality follows from the fixed-point condition when $g_r^\star>0$, and the expression is zero otherwise. Thus $\delta_{g^\star}$ is exactly obedient and belongs to $\mathcal O_k^\epsilon(\policy_{-k})$ for every $\epsilon\geq0$.

    \textit{Closed graph and upper hemicontinuity.} Take sequences $\policy_{-k}^n\to\policy_{-k}$ and $\policy_k^n\to\policy_k$ with $\policy_k^n\in\mathcal O_k^\epsilon(\policy_{-k}^n)$ for every $n$. For every $r,q\in\mathcal R$, Lemma~\ref{lem:policy-continuity} gives
    \begin{gather*}
        \mathbb E_{\policy_k,\policy_{-k}}\bigl[
            f_r^{(k)}\bigl(C_r(f)-C_q(f)-\epsilon\bigr)
        \bigr]\\
        =\lim_{n\to\infty}
        \mathbb E_{\policy_k^n,\policy_{-k}^n}\bigl[
            f_r^{(k)}\bigl(C_r(f)-C_q(f)-\epsilon\bigr)
        \bigr]\leq0.
    \end{gather*}
Therefore $\policy_k\in\mathcal O_k^\epsilon(\policy_{-k})$. Since the policy spaces are metrizable, this shows that the graph is closed. The codomain $\Delta(\mathcal F_k)$ is compact, so the closed-graph characterization implies upper hemicontinuity \citep{border1985fixed,aliprantis2006infinite}.
\end{proof}

\subsection{Existence Theorem}
\foeexistence*

\begin{proof}[Proof of Theorem~\ref{thm:foe-existence}]
For mediator $k$, apply Berge's maximum theorem in minimization form to $J_k$ and $\mathcal O_k^\epsilon$ on $\Phi_{-k}$. The objective is jointly continuous by Lemma~\ref{lem:policy-continuity}. The feasible correspondence has nonempty compact values by Lemma~\ref{lem:obedient-policy-sets} and is continuous by Lemma~\ref{lem:obedient-policy-sets} and Assumption~\ref{ass:obedient-policy-lhc}. Therefore the optimizer correspondence
    \[
        BR_k^\epsilon(\policy_{-k})
        =\operatorname*{arg\,min}_{\policy_k\in\mathcal O_k^\epsilon(\policy_{-k})}
        \bigl[J_k(\policy_k,\policy_{-k})\bigr]
    \]
has nonempty compact values and is upper hemicontinuous. Its values are convex because $\mathcal O_k^\epsilon(\policy_{-k})$ is convex and $J_k(\cdot,\policy_{-k})$ is linear.

Define
    \begin{equation}
        BR^\epsilon:\Phi\rightrightarrows\Phi,
        \text{ with }
        BR^\epsilon(\policy_{1:K})
        :=\prod_{k\in[K]}BR_k^\epsilon(\policy_{-k}).
        \label{eq:joint-best-response}
    \end{equation}
As a finite product of upper-hemicontinuous correspondences with nonempty compact convex values, $BR^\epsilon$ has these same properties.

Apply the Kakutani--Fan--Glicksberg fixed-point theorem to $BR^\epsilon$ on $\Phi$ \citep{delatte2022kakutani}. By Lemma~\ref{lem:policy-spaces}, $\Phi$ is a nonempty compact convex subset of a locally convex Hausdorff topological vector space. The correspondence $BR^\epsilon$ is upper hemicontinuous with nonempty compact convex values. Therefore there exists $\policy_{1:K}^\star\in\Phi$ such that
    \[
        \policy_{1:K}^\star\in BR^\epsilon(\policy_{1:K}^\star).
    \]
Thus $\policy_k^\star\in BR_k^\epsilon(\policy_{-k}^\star)$ for every $k\in[K]$, so $\policy_{1:K}^\star\in\operatorname{FOE}_K(\epsilon)$.
\end{proof}

\section{Proofs for the Moment Formulation}\label{app:moment-proof}

\momentfeasibility*

\begin{proof}[Proof of Proposition~\ref{prop:moment-feasibility}]
Each coordinate of $g$ and $gg^\top$ is bounded and continuous on the compact simplex $\mathcal F(s)$, so the moment map $T_s$ is continuous under weak convergence. It is also affine. The policy space $\Delta(\mathcal F(s))$ is nonempty, compact, and convex by the same argument as Lemma~\ref{lem:policy-spaces}; hence its image under $T_s$ has these properties.

We prove equality with $\mathfrak M_s$ in both directions. Every point of $\mathfrak M_s$ is a finite convex combination $\sum_i p_i(g^i,g^i(g^i)^\top)$, where $g^i\in\mathcal F(s)$, $p_i\geq0$, and $\sum_i p_i=1$. The policy $\sum_i p_i\delta_{g^i}$ has precisely these moments, so $\mathfrak M_s\subseteq T_s(\Delta(\mathcal F(s)))$. Here $\delta_g$ denotes the point mass at $g$.

For the reverse inclusion, the set $\{(g,gg^\top):g\in\mathcal F(s)\}$ is compact as the continuous image of the compact simplex $\mathcal F(s)$. Its finite-dimensional convex hull $\mathfrak M_s$ is also compact by Carath\'{e}odory's theorem. If $T_s(\policy)$ lay outside this compact convex set, the separating-hyperplane theorem would give a linear functional $L$ such that
    \[
        L(T_s(\policy))>\max_{(u,V)\in\mathfrak M_s}L(u,V).
    \]
Yet
    \[
        L(T_s(\policy))=\mathbb E_\policy[L(g,gg^\top)]\leq\max_{g\in\mathcal F(s)}L(g,gg^\top)=\max_{(u,V)\in\mathfrak M_s}L(u,V),
    \]
contradicting strict separation. Hence $T_s(\Delta(\mathcal F(s)))\subseteq\mathfrak M_s$. Combining this inclusion with the finite-mixture construction above shows that every policy, including one with infinite support, has a finitely supported policy with the same first and second moments.

To bound the support size, set $D=R(R+1)/2$. The matrices $gg^\top$, $g\in\mathcal F(s)$, lie in the affine hyperplane $\mathbf1^\top Q\mathbf1=s^2$ of $\mathbb S^R$, of dimension $D-1$, since $\mathbf1^\top gg^\top\mathbf1=(\mathbf1^\top g)^2=s^2$. Carath\'{e}odory's theorem \citep[Theorem~17.1]{rockafellar1970convex} therefore expresses any feasible $Q$ as a convex combination of at most $D$ such matrices. The associated policy also realizes $m$, since every feasible pair satisfies $Q\mathbf1=sm$ by the fixed mass of each flow.
\end{proof}

\begin{restatable}[Necessary properties of feasible moments]{lemma}{momentidentities}\label{lem:moment-identities}
Every $(m,Q)\in\mathfrak M_s$ satisfies $m\geq0$, $\mathbf1^\top m=s$, $Q\mathbf1=sm$, $Q_{rj}\geq0$, $Q\succeq0$, and $Q-mm^\top\succeq0$. If $m_r=0$, then $Q_{rj}=0$ for every $j$. Moreover, $Q=mm^\top$ if and only if every policy realizing $(m,Q)$ selects the single flow $m$ with probability one.
\end{restatable}

\begin{proof}
Take a realizing policy, which exists by Proposition~\ref{prop:moment-feasibility}. Nonnegativity of $g$ gives $m\geq0$ and $Q_{rj}\geq0$. Since $\mathbf1^\top g=s$ almost surely,
    \[
        \mathbf1^\top m=s,\qquad Q\mathbf1=\mathbb E[g(g^\top\mathbf1)]=s\mathbb E[g]=sm.
    \]
For every $u\in\mathbb R^R$,
    \[
        u^\top Qu=\mathbb E[(u^\top g)^2]\geq0,\qquad u^\top(Q-mm^\top)u=\mathbb E[(u^\top(g-m))^2]\geq0.
    \]
If $m_r=0$, then the nonnegative random variable $g_r$ has zero expectation, so $g_r=0$ almost surely and $Q_{rj}=\mathbb E[g_rg_j]=0$. If $Q=mm^\top$, every coordinate has zero variance, so $g=m$ almost surely under any realizing policy. The converse holds for the point mass at $m$.
\end{proof}

These are necessary properties of realizable moments. Proposition~\ref{prop:moment-cp} characterizes realizability exactly using complete positivity; Theorem~\ref{cor:moment-sdp} permits its replacement by double nonnegativity when $R\leq4$.

\momentcostobedience*

\begin{proof}[Proof of Proposition~\ref{prop:moment-cost-obedience}]
To obtain \eqref{eq:moment-cross-flow}, split the aggregate flow as $f_j=f_j^{(k)}+\sum_{\ell\neq k}f_j^{(\ell)}$. Independence across mediators gives
    \[
        \begin{aligned}
            \mathbb E_{\policy_k,\policy_{-k}}[f_r^{(k)}f_j]
            &=\mathbb E_{\policy_k}[f_r^{(k)}f_j^{(k)}]+\sum_{\ell\neq k}\mathbb E_{\policy_k}[f_r^{(k)}]\mathbb E_{\policy_\ell}[f_j^{(\ell)}]\\
            &=Q_{rj}+m_r(v_k)_j.
        \end{aligned}
    \]
Equation~\eqref{eq:moment-affine-costs} gives $C_r(f)=\sum_jH_{rj}f_j+d_r$. Applying \eqref{eq:moment-cross-flow} gives
    \[
        \begin{aligned}
            J_k(\policy_k,\policy_{-k})
            &=\mathbb E\!\left[\sum_r f_r^{(k)} \left(\sum_jH_{rj}f_j+d_r\right) \right]\\
            &=\sum_{r,j}H_{rj}\mathbb E[f_r^{(k)}f_j]+\sum_r d_r\mathbb E[f_r^{(k)}]\\
            &=\sum_{r,j}H_{rj}\bigl(Q_{rj}+m_r(v_k)_j\bigr)+\sum_r d_rm_r\\
            &=\langle H,Q\rangle+m^\top Hv_k+d^\top m.
        \end{aligned}
    \]
Similarly, the unconditional obedience expression equals
    \[
        \begin{aligned}
            &\mathbb E_{\policy_k,\policy_{-k}}\left[f_r^{(k)}\bigl(C_r(f)-C_q(f)-\epsilon\bigr)\right]\\
            &\qquad=\sum_j(H_{rj}-H_{qj})\mathbb E[f_r^{(k)}f_j]+(d_r-d_q-\epsilon)\mathbb E[f_r^{(k)}]\\
            &\qquad=\sum_j(H_{rj}-H_{qj})\bigl(Q_{rj}+m_r(v_k)_j\bigr)+(d_r-d_q-\epsilon)m_r.
        \end{aligned}
    \]
The expansion above shows that \eqref{eq:moment-obedience} is equivalent to the unconditional obedience constraints in \eqref{eq:unconditional-obedience}. Comparing a route with itself imposes no restriction: when $q=r$, the inequality reduces to $-\epsilon m_r\leq0$. A route that is never recommended also imposes no restriction: if $m_r=0$, Lemma~\ref{lem:moment-identities} gives $Q_{rj}=0$ for every $j$, so all corresponding inequalities reduce to $0\leq0$.

At a full policy profile, $\sum_kJ_k(\policy_k,\policy_{-k})=\mathbb E_{\policy_{1:K}}[SC(f)]$. Substituting \eqref{eq:moment-mediator-objective} and using $v_k=\sum_{\ell\neq k}m^{(\ell)}$ gives
    \[
        \mathbb E_{\policy_{1:K}}[SC(f)]=\sum_k\langle H,Q^{(k)}\rangle+\sum_{k\neq\ell}(m^{(k)})^\top Hm^{(\ell)}+d^\top\bar m.
    \]
Symmetry of $H$ combines the ordered cross terms into twice the sum over $k<\ell$, proving \eqref{eq:moment-social-cost}.
\end{proof}

For a set of policies $S\subseteq\Delta(\mathcal F_k)$, write $T_{s_k}(S):=\{T_{s_k}(\policy_k):\policy_k\in S\}$.
\begin{restatable}[Moment feasible and best-response sets]{proposition}{momentbestresponse}\label{prop:moment-best-response}
Fix rival policies $\policy_{-k}$ with aggregate mean $v_k$. Then
    \begin{equation}
        \begin{aligned}
            T_{s_k}\bigl(\mathcal O_k^\epsilon(\policy_{-k})\bigr)&=\mathfrak O_k^\epsilon(v_k),\\
            T_{s_k}\bigl(BR_k^\epsilon(\policy_{-k})\bigr)&=BR_{k,\mathrm{mom}}^\epsilon(v_k).
        \end{aligned}
        \label{eq:moment-best-response-image}
    \end{equation}
For every candidate policy $\policy_k\in\Delta(\mathcal F_k)$,
    \begin{equation}
        \policy_k\in BR_k^\epsilon(\policy_{-k})
        \quad\Longleftrightarrow\quad
        T_{s_k}(\policy_k)\in BR_{k,\mathrm{mom}}^\epsilon(v_k).
        \label{eq:moment-best-response-equivalence}
    \end{equation}
The moment feasible and best-response sets are nonempty, compact, and convex.
\end{restatable}

\begin{proof}
    \textit{Obedient sets.} By Proposition~\ref{prop:moment-cost-obedience}, every policy obedient against $\policy_{-k}$ has moments in $\mathfrak O_k^\epsilon(v_k)$. Conversely, every pair in $\mathfrak O_k^\epsilon(v_k)$ has a realizing policy by Proposition~\ref{prop:moment-feasibility}. Any such policy is obedient because its moments satisfy \eqref{eq:moment-obedience}. Thus $T_{s_k}(\mathcal O_k^\epsilon(\policy_{-k}))=\mathfrak O_k^\epsilon(v_k)$, proving the first equality in \eqref{eq:moment-best-response-image}.

    \textit{Set properties.} The obedient policy set is nonempty by Lemma~\ref{lem:obedient-policy-sets}, so the first equality implies that $\mathfrak O_k^\epsilon(v_k)$ is nonempty. For fixed $v_k$, the obedience constraints in \eqref{eq:moment-obedience} are linear inequalities in $(m,Q)$. They therefore define a closed convex subset of the compact convex set $\mathfrak M_{s_k}$. Hence $\mathfrak O_k^\epsilon(v_k)$ is compact and convex. The continuous objective in \eqref{eq:moment-best-response} attains its minimum on this nonempty compact set, and its minimizer set is compact. Linearity of the objective also makes the minimizer set convex.

    \textit{Best responses.} Every obedient policy has a feasible obedient moment pair, and every feasible obedient moment pair is realized by an obedient policy. By \eqref{eq:moment-mediator-objective}, the policy and its moment pair have the same objective value. Minimizing over obedient policies is therefore equivalent to minimizing over obedient moment pairs. Thus a policy is a best response if and only if its moments are a moment best response, proving \eqref{eq:moment-best-response-equivalence}. Since every moment best response has a realizing policy, this also proves the second equality in \eqref{eq:moment-best-response-image}.
\end{proof}

\momentequivalence*

\begin{proof}[Proof of Theorem~\ref{thm:moment-equivalence}]
For a policy profile $\policy_{1:K}$, let $(m^{(k)},Q^{(k)})=T_{s_k}(\policy_k)$ and $v_k=\sum_{\ell\neq k}m^{(\ell)}$. By Proposition~\ref{prop:moment-best-response}, $\policy_k\in BR_k^\epsilon(\policy_{-k})$ if and only if $(m^{(k)},Q^{(k)})\in BR_{k,\mathrm{mom}}^\epsilon(v_k)$. Requiring these conditions for every $k$ gives membership in $\operatorname{FOE}_K(\epsilon)$ and $\operatorname{mFOE}_K(\epsilon)$, respectively, proving the equivalence.

Given a profile $((m^{(k)},Q^{(k)}))_{k\in[K]}\in\operatorname{mFOE}_K(\epsilon)$, Proposition~\ref{prop:moment-feasibility} supplies finitely supported policies $\policy_k$ with $T_{s_k}(\policy_k)=(m^{(k)},Q^{(k)})$. Choose any such realizing policies $\policy_k$ and let each mediator independently sample its group flow $f^{(k)}\sim\policy_k$, as required by the model. Their moment profile is unchanged, so the equivalence just proved gives $\policy_{1:K}\in\operatorname{FOE}_K(\epsilon)$.
\end{proof}

\subsection{Consequences of the Moment Cost Formulas}\label{app:moment-cost-consequences}

\paragraph{Conditional expected route-cost difference.} Fix $\policy_k$ with moments $(m,Q)$ and rival policies $\policy_{-k}$ with aggregate mean $v_k$. For $m_r>0$, Lemma~\ref{lem:recommendation-conditioning} gives
\[
    \mathbb E_{\policy_k,\policy_{-k}}[C_r(f)-C_q(f)\mid\operatorname{rec}_k=r]
    =\frac{\mathbb E_{\policy_k,\policy_{-k}}[f_r^{(k)}(C_r(f)-C_q(f))]}{m_r}.
\]
Substituting the affine route costs and \eqref{eq:moment-cross-flow} yields
\begin{equation}
    \begin{aligned}
        &\mathbb E_{\policy_k,\policy_{-k}}[C_r(f)-C_q(f)\mid\operatorname{rec}_k=r]\\
        &\qquad=\sum_{j\in\mathcal R}(H_{rj}-H_{qj})\left(\frac{Q_{rj}}{m_r}+(v_k)_j\right)+d_r-d_q.
    \end{aligned}
    \label{eq:moment-conditional-cost}
\end{equation}
Here $Q_{rj}/m_r$ is the conditional expected own-group flow on route $j$ after recommendation $r$. Independent mediator sampling leaves the conditional expected rival flow equal to $(v_k)_j$. If $m_r=0$, the recommendation has probability zero and no conditional expectation is required.

\paragraph{Expected social cost and covariance.} Fix a policy profile $\policy_{1:K}$ and write $\bar m:=\mathbb E[f]=\sum_km^{(k)}$. Define the centered aggregate flow $\xi:=f-\bar m$, so $f=\bar m+\xi$ and $\mathbb E[\xi]=0$. By definition,
\[
    \operatorname{Cov}(f)
    =\mathbb E\!\left[(f-\mathbb E[f])(f-\mathbb E[f])^\top\right]
    =\mathbb E[\xi\xi^\top].
\]
Equation~\eqref{eq:moment-flow-social-cost} gives $SC(f)=f^\top Hf+d^\top f$. Substituting $f=\bar m+\xi$ and taking expectations gives
\[
    \begin{aligned}
        \mathbb E[SC(f)]
        &=\mathbb E\!\left[(\bar m+\xi)^\top H(\bar m+\xi)+d^\top(\bar m+\xi)\right]\\
        &=\bar m^\top H\bar m+\bar m^\top H\mathbb E[\xi]
          +\mathbb E[\xi]^\top H\bar m\\
        &\qquad+\mathbb E[\xi^\top H\xi]+d^\top\bar m+d^\top\mathbb E[\xi]\\
        &=SC(\bar m)+\mathbb E[\xi^\top H\xi].
    \end{aligned}
\]
The terms linear in $\xi$ vanish because $\mathbb E[\xi]=0$. Using $\xi^\top H\xi=\operatorname{tr}(H\xi\xi^\top)$, the remaining term is
\[
    \mathbb E[\xi^\top H\xi]
    =\operatorname{tr}\!\left(H\mathbb E[\xi\xi^\top]\right)
    =\operatorname{tr}\!\left(H\operatorname{Cov}(f)\right).
\]
For each mediator $k$, its group-flow covariance is
\[
    \begin{aligned}
        \Sigma_k:=\operatorname{Cov}(f^{(k)})
        &=\mathbb E\!\left[f^{(k)}(f^{(k)})^\top\right]
          -\mathbb E[f^{(k)}]\mathbb E[f^{(k)}]^\top\\
        &=Q^{(k)}-m^{(k)}(m^{(k)})^\top.
    \end{aligned}
\]
Since $f=\sum_kf^{(k)}$ and mediator samples are independent, cross-group covariances vanish. Thus $\operatorname{Cov}(f)=\sum_k\Sigma_k$, and the decomposition becomes
\begin{equation}
    \begin{aligned}
        \mathbb E_{\policy_{1:K}}[SC(f)]
        &=SC(\bar m)+\sum_k\operatorname{tr}(H\Sigma_k)\\
        &=SC(\bar m)+\sum_k\operatorname{tr}\!\left(H\bigl[Q^{(k)}-m^{(k)}(m^{(k)})^\top\bigr]\right).
    \end{aligned}
    \label{eq:moment-social-cost-covariance}
\end{equation}
This separates social cost at the mean flow from the contribution of random fluctuations. Since $H\succeq0$, $\xi^\top H\xi\geq0$, so $\mathbb E[SC(f)]\geq SC(\bar m)$. This comparison holds at fixed mean flows; deterministic mean-flow policies need not satisfy obedience.

\subsection{Completely Positive Representation and Semidefinite Best Responses}\label{app:moment-conic}

Lemma~\ref{lem:moment-identities} gives necessary properties of realizable moments, but these alone do not guarantee a realizing policy when $R>4$. We now show that complete positivity together with the mass equalities characterizes realizability exactly. For $R\leq4$, this characterization yields the semidefinite formulation.

\begin{restatable}[Completely positive matrix]{definition}{completelypositivedefinition}\label{def:moment-cp}
A symmetric matrix $Q\in\mathbb S^R$ is \emph{completely positive} if it admits a finite decomposition
    \[
        Q=\sum_{i=1}^N u^i(u^i)^\top,\qquad u^i\in\mathbb R_+^R.
    \]
Denote the cone of these matrices by $\mathrm{CP}_R$.
\end{restatable}

\begin{restatable}[Completely positive representation of feasible moments]{proposition}{momentcp}\label{prop:moment-cp}
For every $s>0$ and finite route count $R$,
    \begin{equation}
        \mathfrak M_s=\left\{(m,Q)\in\mathbb R^R\times\mathbb S^R:
        Q\in\mathrm{CP}_R,\quad Q\mathbf1=sm,\quad\mathbf1^\top m=s\right\}.
        \label{eq:moment-cp}
    \end{equation}
\end{restatable}

\begin{proof}[Proof of Proposition~\ref{prop:moment-cp}]
If $(m,Q)\in\mathfrak M_s$, Proposition~\ref{prop:moment-feasibility} supplies a finite representation $m=\sum_i p_ig^i$ and $Q=\sum_i p_ig^i(g^i)^\top$, where $g^i\in\mathcal F(s)$, $p_i\geq0$, and $\sum_i p_i=1$. Taking $u^i=\sqrt{p_i}g^i$ proves $Q\in\mathrm{CP}_R$. The mass equalities $Q\mathbf1=sm$ and $\mathbf1^\top m=s$ follow from Lemma~\ref{lem:moment-identities}.

Conversely, suppose $Q\in\mathrm{CP}_R$, $Q\mathbf1=sm$, and $\mathbf1^\top m=s$. Choose a decomposition $Q=\sum_i u^i(u^i)^\top$ with $u^i\geq0$ and discard zero factors. For each remaining factor, define
    \begin{equation}
        t_i:=\mathbf1^\top u^i>0,\qquad
        p_i:=\frac{t_i^2}{s^2},\qquad
        g^i:=\frac{s}{t_i}u^i.
        \label{eq:moment-cp-recovery}
    \end{equation}
Each $g^i$ is nonnegative with total mass $s$. Moreover,
    \[
        \sum_i p_i=\frac{\sum_i t_i^2}{s^2}
        =\frac{\mathbf1^\top Q\mathbf1}{s^2}
        =\frac{s\mathbf1^\top m}{s^2}=1.
    \]
Thus $\policy:=\sum_i p_i\delta_{g^i}$ is a feasible policy. Its moments are
    \[
        \begin{aligned}
            \sum_i p_i g^i&=\frac1s\sum_i t_i u^i=\frac{Q\mathbf1}{s}=m,\\
            \sum_i p_i g^i(g^i)^\top&=\sum_i u^i(u^i)^\top=Q.
        \end{aligned}
    \]
Hence $(m,Q)\in\mathfrak M_s$, proving the reverse inclusion. Equation~\eqref{eq:moment-cp-recovery} also gives an explicit realizing policy once nonnegative factors are available.
\end{proof}

\begin{restatable}[Doubly nonnegative matrix]{definition}{doublynonnegativedefinition}\label{def:moment-dnn}
A matrix is \emph{doubly nonnegative} if it is positive semidefinite and entrywise nonnegative. Write
\[
    \mathrm{DNN}_R:=\{Q\in\mathbb S^R:Q\succeq0,\ Q_{rj}\geq0\text{ for all }r,j\}.
\]
\end{restatable}

The matrix properties derived directly from realizing policies in Lemma~\ref{lem:moment-identities} therefore say that $Q\in\mathrm{DNN}_R$. Unlike complete positivity together with the mass equalities, double nonnegativity and those equalities need not guarantee a realizing policy when $R>4$. For smaller matrices, however, the two cones coincide.

Every completely positive matrix is doubly nonnegative. The converse holds exactly through matrix order four: $\mathrm{CP}_R=\mathrm{DNN}_R$ for $R\leq4$, whereas $\mathrm{CP}_R\subsetneq\mathrm{DNN}_R$ for $R > 4$ \citep[Proposition~8.6]{burer2012copositive}.

\momentsdp*

\begin{proof}[Proof of Theorem~\ref{cor:moment-sdp}]
For $R\leq4$, substitute $\mathrm{CP}_R=\mathrm{DNN}_R$ into \eqref{eq:moment-cp}. This gives the exact feasible moment set through positive semidefiniteness, entrywise nonnegativity, and the two mass equalities. Adding the obedience inequalities and minimizing the moment objective gives \eqref{eq:moment-sdp}. With $v_k$ fixed, the objective and scalar constraints are linear, so this is an SDP. Proposition~\ref{prop:moment-best-response} identifies its solutions with policy best responses through the moment map.
\end{proof}

\clearpage
\section{Numerical Details}\label{app:numerics}

\subsection{Search and Recovery Settings}\label{app:numerical-method}

The illustration uses sequential updates with $\lambda=0.5$ and $T=160$. Randomized starts use seeds 0 and 1, $R+1$ allocations from $s_k\operatorname{Dirichlet}(\mathbf1)$, and Dirichlet mixture weights. The Wardrop start is $g_k=s_kw$. These profiles are realizable but need not be obedient. No objective perturbation or tie-breaking term is added.

CVXPY passes normalized variables $u=m/s_k$, $Z=Q/s_k^2$ to Clarabel. SCS crosschecks use at most 100,000 iterations. Recovery tries deterministic, diagonal, and spectral factors before bounded nonnegative least squares, with at most twelve randomized starts, 2,000 function evaluations per start, and $R(R+1)/2$ factor columns. Unused routes are omitted and diagonal scaling is applied during fitting. The function descriptions below give the checks and numerical thresholds.

\subsection{Verification and Thresholds}\label{app:numerical-residuals}

The checks are grouped by the functions in Algorithm~\ref{alg:numerical-procedure}. Numerical acceptance thresholds are distinct from the behavioral tolerance $\epsilon$. Requested solver accuracies, check triggers, and recovery cutoffs serve different purposes and do not themselves establish equilibrium.

\paragraph{\textnormal{\textsc{ShouldCheck}}.} This function schedules verification initially, every five sweeps, at the iteration limit, or when the normalized iterate change is small. For $t\geq1$, define
\[
    \Delta^t=\max_k\max\left\{
    \frac{\|m_k^t-m_k^{t-1}\|_\infty}{s_k},
    \frac{\|Q_k^t-Q_k^{t-1}\|_\infty}{s_k^2}\right\}.
\]
The infinity norm denotes the largest absolute entry. A small update triggers verification rather than acceptance.

\begin{table}[H]
\centering\small\setlength{\tabcolsep}{4pt}
\caption{\textsc{ShouldCheck}: additional check trigger.}\label{tab:numerical-check-trigger}
\begin{tabular}{>{\raggedright\arraybackslash}p{.49\linewidth}>{\raggedright\arraybackslash}p{.12\linewidth}>{\raggedright\arraybackslash}p{.32\linewidth}}
\toprule
Quantity & Threshold & Description\\
\midrule
$\Delta^t$ & $\leq10^{-6}$ & Normalized iterate change\\
\bottomrule
\end{tabular}
\end{table}

\paragraph{\textnormal{\textsc{SolveBestResponse}}.} This function solves the moment SDP against fixed rival means. The returned moments must pass feasibility and obedience checks regardless of solver status. For an \texttt{optimal\_inaccurate} result used in equilibrium verification, an SCS crosscheck must also agree with the Clarabel objective. Write $\widehat b_k^{\mathrm C}$ and $\widehat b_k^{\mathrm S}$ for these two values against the same rivals, and $\mathcal L_{krq}(x)$ for the left-hand side of \eqref{eq:moment-obedience}.

For each group, let $u=m/s_k$, $Z=Q/s_k^2$, $Z_{\mathrm{sym}}=(Z+Z^\top)/2$, and $[z]_+=\max\{z,0\}$. The infinity norm below denotes the largest absolute entry. The moment residual combines raw and normalized checks:
\begin{equation}
    \begin{aligned}
        E_{\mathrm{mom}}(m,Q;s_k):=\max\bigl\{&
        |\mathbf1^\top m-s_k|,\;\|Q\mathbf1-s_km\|_\infty,\\
        &|\mathbf1^\top u-1|,\;\|Z\mathbf1-u\|_\infty,\;\|Z-Z^\top\|_\infty,\\
        &[-\min_r u_r]_+,\;[-\min_{r,j}Z_{rj}]_+,\\
        &[-\lambda_{\min}(Z_{\mathrm{sym}})]_+,\;[-\lambda_{\min}(Z_{\mathrm{sym}}-uu^\top)]_+\bigr\}.
    \end{aligned}
    \label{eq:numerical-moment-residuals}
\end{equation}

\begin{table}[H]
\centering\small\setlength{\tabcolsep}{4pt}
\caption{\textsc{SolveBestResponse}: solver settings and solution checks.}\label{tab:numerical-solver-checks}
\begin{tabular}{>{\raggedright\arraybackslash}p{.49\linewidth}>{\raggedright\arraybackslash}p{.12\linewidth}>{\raggedright\arraybackslash}p{.32\linewidth}}
\toprule
Quantity or setting & Limit & Description\\
\midrule
Clarabel $\mathrm{tol}_{\mathrm{gap,abs}}$, $\mathrm{tol}_{\mathrm{gap,rel}}$, $\mathrm{tol}_{\mathrm{feas}}$ & $10^{-9}$ & Gap and feasibility accuracy\\
SCS $\mathrm{eps}_{\mathrm{SCS}}$ & $10^{-9}$ & Solver accuracy\\
$E_{\mathrm{mom}}(m,Q;s_k)$ & $\leq10^{-6}$ & Moment feasibility\\
$\max_{r\neq q}[\mathcal L_{krq}(x)]_+$ & $\leq10^{-6}$ & User obedience constraint\\
$|\widehat b_k^{\mathrm C}-\widehat b_k^{\mathrm S}|$, when required & $\leq10^{-6}$ & Agreement between solver objectives\\
\bottomrule
\end{tabular}
\end{table}

\paragraph{\textnormal{\textsc{VerifyMomentEquilibrium}}.} For a candidate profile $x$, this function checks each mediator's moment feasibility, obedience, and optimality against the current rival means $v_k=\sum_{\ell\neq k}m^{(\ell)}$. Let $j_k(x_k;v_k)$ be the objective in \eqref{eq:moment-mediator-objective}. A fresh \textsc{SolveBestResponse} call gives $\widehat b_k(v_k)$, and the computed best-response gap is
\begin{equation}
    G_k(x):=j_k(x_k;v_k)-\widehat b_k(v_k).
    \label{eq:numerical-br-gap}
\end{equation}
This gap measures the available cost reduction against fixed rivals, rather than the change in cost between sweeps. Acceptance requires every check below to pass.

\begin{table}[H]
\centering\small\setlength{\tabcolsep}{4pt}
\caption{\textsc{VerifyMomentEquilibrium}: candidate-profile checks.}\label{tab:numerical-verification}
\begin{tabular}{>{\raggedright\arraybackslash}p{.49\linewidth}>{\raggedright\arraybackslash}p{.12\linewidth}>{\raggedright\arraybackslash}p{.32\linewidth}}
\toprule
Quantity & Threshold & Description\\
\midrule
$\max_k E_{\mathrm{mom}}(m^{(k)},Q^{(k)};s_k)$ & $\leq10^{-6}$ & Moment feasibility\\
$\max_{k,r\neq q}[\mathcal L_{krq}(x)]_+$ & $\leq10^{-6}$ & User obedience constraint\\
$\max_k|G_k(x)|$ & $\leq10^{-6}$ & Mediator best-response optimality\\
\bottomrule
\end{tabular}
\end{table}

\paragraph{\textnormal{\textsc{RealizePolicies}}.} Using the construction in \eqref{eq:moment-cp-recovery} with the normalized factorization $BB^\top\approx Z$, let $b^i$ denote column $i$ of $B$ and $\beta_i=\mathbf1^\top b^i$. After discarding columns with $\beta_i\leq10^{-14}$, form $g^i=s_kb^i/\beta_i$ with probabilities proportional to $\beta_i^2$. Each retained column yields one flow allocation, so the recovered policy may use fewer than $R(R+1)/2$ allocations. The function checks the factorization, allocations, probabilities, and reconstruction of the target moments. A failed factor search is recorded rather than accepted. Write
\[
    \widetilde m=\sum_i p_ig^i,\qquad
    \widetilde Q=\sum_i p_ig^i(g^i)^\top,
\]
and define
\[
    E_{\mathrm{pol}}:=\max\left\{
    \max_i|\mathbf1^\top g^i-s_k|,\;
    \left|\sum_i p_i-1\right|,\;
    [-\min_{i,r}g_r^i]_+,\;
    [-\min_i p_i]_+
    \right\}.
\]
The tildes distinguish reconstructed moments from the target moments. The recovered profile $\policy_{1:K}$ must also pass \textsc{VerifyMomentEquilibrium}, using its reconstructed moments and rival means. We evaluate $J_k(\policy_k,\policy_{-k})$ directly from the recovered allocations and compare it with $j_k(x_k;v_k)$. For social cost, we compare $\sum_k j_k(x_k;v_k)$ with the independent physical-edge calculation
\begin{equation}
    S_{\mathrm{edge}}=\sum_e\left[a_e\left((\mathbb E[\edgeflow_e])^2+\sum_k\operatorname{Var}(\edgeflow_e^{(k)})\right)+b_e\mathbb E[\edgeflow_e]\right],
    \label{eq:numerical-edge-check}
\end{equation}
where edge-load means and variances are computed from the recovered policies.

\begin{table}[H]
\centering\small\setlength{\tabcolsep}{4pt}
\caption{\textsc{RealizePolicies}: reconstruction checks.}\label{tab:numerical-recovery-checks}
\begin{tabular}{>{\raggedright\arraybackslash}p{.49\linewidth}>{\raggedright\arraybackslash}p{.12\linewidth}>{\raggedright\arraybackslash}p{.32\linewidth}}
\toprule
Quantity & Threshold & Description\\
\midrule
$\|\widetilde m-m\|_\infty$, $\|\widetilde Q-Q\|_\infty$ & $\leq10^{-6}$ & Moment reconstruction\\
$\|\widetilde m/s_k-u\|_\infty$, $\|\widetilde Q/s_k^2-Z\|_\infty$ & $\leq10^{-6}$ & Normalized moment reconstruction\\
$E_{\mathrm{pol}}$ & $\leq10^{-6}$ & Allocation and probability feasibility\\
$\|BB^\top-Z\|_\infty$ & $\leq10^{-6}$ & Factorization error\\
$\max_k|J_k(\policy_k,\policy_{-k})-j_k(x_k;v_k)|$ & $\leq10^{-6}$ & Mediator-cost consistency\\
$|S_{\mathrm{edge}}-\sum_k j_k(x_k;v_k)|$ & $\leq10^{-6}$ & Social-cost consistency\\
\bottomrule
\end{tabular}
\end{table}

All reconstruction checks apply to each mediator where appropriate. A small flow-weighted obedience residual does not bound conditional saving errors by the same amount on rarely recommended routes. Solver status or a small iterate change alone does not establish acceptance.

\subsection{Benchmark Justification}\label{app:numerical-benchmarks}

SO and Wardrop are computed independently by enumerating all nonempty route supports, solving their linear first-order equations, and checking feasibility and active and inactive gradients. Pooling uses the mass-one SDP with $v=0$, followed by recovery and verification.

\begin{proof}[Justification of \eqref{eq:numerical-pooling-bound}]
Let $\policy_{1:K}$ be an obedient fragmented profile and let $\policy$ be the distribution of $f=\sum_kf^{(k)}$. Every realization belongs to $\mathcal F(1)$. For each $r,q$, summing the group obedience inequalities gives
\[
    \mathbb E_\policy[f_r(C_r(f)-C_q(f)-\epsilon)]
    =\sum_k\mathbb E_{\policy_{1:K}}[f_r^{(k)}(C_r(f)-C_q(f)-\epsilon)]\leq0.
\]
Hence $\policy$ is feasible for the pooled problem and has the same expected social cost. This proves $P(\epsilon)\leq\mathbb E[SC(f)]$. Every aggregate realization has cost at least $SC(f^{\mathrm{SO}})$, proving the other inequality.
\end{proof}

At unit demand, the maximum possible route-cost saving is
\begin{equation}
    \epsilon_{\max}=\max_{r,q}\left\{d_r-d_q+\max_j(H_{rj}-H_{qj})\right\}.
    \label{eq:numerical-epsilon-max}
\end{equation}
The maximum of the affine difference $C_r(f)-C_q(f)$ on the simplex occurs at a vertex. Including $r=q$ makes this expression nonnegative. For $\epsilon\geq\epsilon_{\max}$, every realization satisfies every route-deviation inequality, so all policies are obedient. This explains the endpoint used in the illustration in Section~\ref{sec:numerical-design}.

\subsection{Configuration and Reproducibility Record}\label{app:numerical-inventory}

The network in Section~\ref{sec:numerical-design} has route order $(T,M,B)$ and affine route-cost parameters
\[
    H=\begin{pmatrix}1&1&0\\1&2&1\\0&1&1\end{pmatrix},\qquad
    d=\begin{pmatrix}0.75\\0\\0.75\end{pmatrix}.
\]
We use $K=2,4,8$ with $s_k=1/K$ and the tolerance grid
\[
    \epsilon\in\{0,0.025,0.05,0.075,0.1,0.15,0.2,0.25,0.3,1\}.
\]
Here $\epsilon_{\max}=1$ in \eqref{eq:numerical-epsilon-max}, so the final point has unrestricted obedience. The SO flow is $(0.5,0,0.5)$; the Wardrop flow used for initialization is $(0.25,0.5,0.25)$.

For each population partition and tolerance, we retain the two randomized starts and the proportional Wardrop start described in Appendix~\ref{app:numerical-method}. The three choices $K\in\{2,4,8\}$, ten tolerance values, and three initializations per configuration give $3\times10\times3=90$ runs. All 90 runs pass the recovered-policy verification checks. We take the median cost across the three starts and retain their minimum and maximum for the bands.

\end{document}